# Observation of unusual optical band structure of $CH_3NH_3PbI_3$ perovskite single crystal

Wei Huang[1,2,†], Shizhong Yue[1,2,†], Yu Liu[1,2], Laipan Zhu[1,3], Peng Jin[1,2], Qing Wu[1,2], Yang Zhang[1,2], Yanan Chen[1,2], Kong Liu[1,2], Ping Liang[1], Shengchun Qu[1,2], Zhijie Wang[1,2]* and Yonghai Chen[1,2]*

[1] Key Laboratory of Semiconductor Materials Science, Beijing Key Laboratory of Low Dimensional Semiconductor Materials and Devices, Institute of Semiconductors, Chinese Academy of Sciences, Beijing, 100083, P. R. China.

[2] College of Materials Science and Opto-Electronic Technology, University of Chinese Academy of Sciences, Beijing, 100049, P. R. China.

[3] Beijing Institute of Nanoenergy and Nanosystems, Chinese Academy of Sciences, Beijing, 100083, P. R. China.

# ABSTRACT

Extensive efforts have been undertaken on the photoelectric physics of hybrid organolead halide perovskites to unveil the reason for the attractive photovoltaic performance. Yet, the resulting evidence is far from being fully conclusive. Herein, we provide another direct support for this issue. In addition to the observation on the conventional band edge at 1.58 eV that presents a blueshift toward temperature increase, interestingly, we also observe an unusual optical band edge at 1.48 eV in $CH_3NH_3PbI_3$ perovskite single crystals for the first time. Contrary to the conventional band edge, this one shows an obvious redshift toward the enhancement in temperature, in agreement with the Varshni relation. More interestingly, the unusual band edge exhibits a series of obvious absorption and photocurrent signals, but the according photoluminescence signals are not observable. This indicates that this band edge is particularly beneficial for the photovoltaic effect due to the inhibited radiative recombination. The kinetics on photo-involved charge transition and transfer are investigated using the pump-probe photoconductivity technique, and a changeable band structure model was proposed.

Hybrid organolead halide perovskites, MAPbX$_3$ (MA=CH$_3$NH$_3^+$, X = Cl, Br, I), have recently received worldwide attention as highly promising materials in solar cells, light-emitting diodes, photodetectors and lasers, attributing to the impressive characteristics including high absorption coefficient, long carrier lifetime and high balanced charge mobility.[1, 2, 3, 4] In particular, the efficiency of solar cells based on hybrid halide perovskites has been realized over 20% certified. Significantly, these devices are usually fabricated on the basis of a solution-processed procedure and the resulting perovskite films are generally polycrystalline.[5, 6, 7] This gives rise to a fundamental question: why such materials are so special in comparison with traditional semiconductors?

To address this issue, extensive efforts have been made for thoroughly elucidating the fundamental photophysics in MAPbI$_3$. Two main controversies, however, are still not well addressed yet. First, what is the real optical band gap (E$_g$) value of MAPbI$_3$ single crystal (MSC)? Some researchers reported that the E$_g$ of MSC was 1.51 eV (concluded from photoluminescence spectroscopy, PL)[8, 9] and 1.48 eV (from diffuse reflectance spectroscopy, DR).[10] However, more reported PL peak positions of MSC are located at 1.61 eV.[11, 12] This indicates that unlike in conventional semiconductors even the E$_g$, the most important parameter of a semiconductor, is not yet definitely determined in MSC. Second, the conventional consensus that MAPbI$_3$ behaves as a direct band gap semiconductor [13, 14] has to be modified, since recent theoretical calculations suggest that the relevant conduction band minimum is slightly shifted in k-space with respect to the valence band maximum, leading to an indirect fundamental band gap.[15, 16, 17] Then Hutter *et al.* have proposed that the band gap in MAPbI$_3$ has a direct-indirect character.[18] These controversies lead to another key question: what does the real band structure of MAPbI$_3$ look like? This is of great significance to the dynamics of both generation and recombination of photo-generated charge carriers, which are far from being well clarified.

The combination of different optical techniques is more effective for a thorough understanding of optical and electronic properties of MAPbI$_3$ perovskite. Yamada, Kanemitsu et al. have determined the band gap energy of MAPbI$_3$ using PL, PC, DR, and other techniques in 2014. [13] This work is of significance to understand the energy band structures of perovskite. Herein, we use a modified PC-R system, the schematic of our PC-R system setup, is shown in **Figure S1**. Through this system, we could realize the in situ optical measurements for both photoconductivity (PC) and reflectance (R) spectra at the same time. This guarantees the accuracy of the results and enables us to observe the unusual optical band structure that have never been reported. We focus on the previously mentioned controversies and investigate the band structure of MSC using techniques based on temperature-dependent PC-R and PL spectroscopies, and pump-probe technique. Surprisingly, we observe an unusual band edge in MSC, which appears obviously in PC and R spectra while not in the relevant PL spectra. This indicates a high conversion efficiency of photon to electric. Details on the origin of this band structure is discussed herein.

■ **EXPERIMENTAL METHODS**

**Synthesis and characterization of MSCs.** MSCs as large as 6 ×5 ×3 mm$^3$ were grown by a solution crystallization process. Equimolar mixture of the CH$_3$NH$_3$I and PbI$_2$ were dissolved in GBA (1, 4-Butyrolactone) for MAPbI$_3$. The concentration of MAPbI$_3$ was 1.2 M. The solution was stirred under 60 ℃ until it turned transparent. Then the temperature increased with the speed of 5 ℃ per 30 min. We maintained the current temperature for 12 h till little black crystals were separated out. As the seed was put into fresh solution, heated and kept at the same temperature for one day, the original seed grew into a larger one. Using the larger crystal as the new seed and repeating the above process, a much larger crystal was obtained. The size of crystal increases in proportion to the repeating times. Then, the MSC was thinned and polished along one direction to 1.5 mm thickness using a mechanical method, in order to control the temperature of the crystal surface faster and more accurately. Finally, a pair of Au electrodes (100 nm of thickness, 1 mm of diameter) with a distance of 2.5 mm were deposited on the top surface by thermal evaporation under high vacuum through a shadow mask. The I-V curves were measured using a digital sourcemeter (Keithley 2450). The XRD patterns were collected using a Rigaku (Dmax 2500) X-ray diffractometer equipped with a Cu Kα X-ray ($\lambda = 1.54186$ Å) tubes operated at 40 kV and 200 mA. Considering the sensitivity of MSCs towards humidity, we installed the MSC in an optical cryostat under high vacuum ($10^{-5}$ Torr) for the series of measurements. We found that vacuum can effectively protect samples from degradation as a result of moisture and other environmental factors (**Figure S2**).

**Temperature-dependent Vis-NIR PC and R spectroscopy.** For optical interband excitation, a supercontinuum laser source (Fianium, WhiteLase™ SC450-4) combined with a monochromator is used, providing excitation source with wavelengths in the range of 450 nm to 1750 nm. The supercontinuum laser provides 5 ps pulses with a repetition rate of 40 MHz and an average power of 4 W. After the monochromatic light with a linewidth of 1.5 nm goes through a chopper, the Gaussian profile light spot with a diameter of about 2 mm irradiates at the central line between two electrodes with a power of about 0.3 mW at 840 nm. The photoconductivity signals, PC, are obtained by applying a dc-electric field between the two electrodes and detecting the PC via a preamplifier and then is recorded by a lock-in amplifier in phase with a mechanically chopper (the reference frequency is 220 Hz). Meanwhile the reflected beam from the sample, after passing through the optical lens, is directed to be focused onto a silicon detector. The reflectivity signals, R, are detected by another lock-in amplifier which is referenced by the same chopper at the same time. A highly refined silver mirror was used as the standard (average reflectance is about 95%) to generated reflectance spectrum. To obtain the temperature-dependent photoconductivity and reflectance spectra, the sample was mounted into an optical cryostat ($10^{-5}$ Torr), allowing the variation of the temperature in the range of 78-350 K. We found that vacuum can effectively protect samples from degradation as a result of moisture and other environmental factors.

**Temperature-dependent PL spectroscopy.** For variable-temperature continuous wave (CW) PL measurements, a frequency-doubled mode-locked Ti: Sapphire laser (440 nm, 4.5 mW) was applied to serve as excitation source. The sample was mounted on a cold finger of a closed-cycle helium refrigerator ($10^{-7}$ Torr) and the sample temperature could be controlled in the range of 8-350 K. The PL was dispersed by a vacuum monochromator and detected by a photomultiplier tube for CW measurements.

**Pump-probe PC spectroscopy**. The pump-probe PC spectroscopy was set up by adding a pump light on the base of the Vis-NIR PC spectroscopy. The changes of PC response of the sample were induced by a photon energy dependent pump light. The pump light was carried out by employing a Ti:sapphire laser with spectral width of 10 nm and a repetition rate of 80 MHz. The excitation wavelength of the pump beam was continually tuned from 680 nm to 1080 nm. The sample is also mounted in an optical cryostat under vacuum ($10^{-5}$ Torr), allowing the variation of the temperature.

## ■ RESULTS AND DISCUSSIONS

**Characterization of MSCs.** First, we used a solution-processed method to obtain MSCs and the according X-ray diffraction (XRD) pattern is given in **Figure 1a** (black curve), where two main peaks from the same crystal plane (200) and (400) are presented. This indicates that MSC orients mainly along (100) direction parallel to the substrate.[9,19] Powder XRD pattern of the ground crystals (red curve) demonstrates a pure perovskite phase for MSC. This photograph of MSC indicates the single crystal nature of the resultant and the according configuration is dodecahedron (inset of **Figure 1a**), a typical crystal habit of a body-centered tetragonal lattice, in agreement with the reported structure at room temperature (space group *I4/mcm*).[20] The surface properties of MSC is shown in **Figure S3**. We can conclude that the MSCs have no other impurity elements and the average roughness of surface is about 70 nm from the X-ray Photoelectron Spectroscopy (XPS) and Atomic Force Microscope (AFM).

Normalized PC (black curve, the electric field is 32 V/cm) and PL spectra (red curve) of the MSC with the incident light perpendicular to sample ($\theta = 0°$) at room temperature are given in **Figure 1b**. The incident angle ($\theta$) of excitation light is defined as the angle between the incident light and the out-of-plane direction of the sample. The location of the PL peak is around 1.58 eV and agrees well with the latest experimental results.[21] The PL spectrum displays only one peak while the PC spectrum presents two peaks. Surprisingly, we find a strong $PC_L$ located at 1.48 eV which is far away from the PL peak, indicating that the $PC_L$ signal is not from the separation of excitons whose recombination would induce a PL emission. This $PC_L$ has never been reported, even though the MSCs have been investigated using temperature-dependent photocurrent spectroscopy. In fact, there indeed is a weak low-energy PC peak which has not ever been noticed by Yoshihiko Kanemitsu et al.. [22] We observed that the light spot on the electrodes in experiments has a great influence on PC spectra (**Figure S4**). And two rectangular-shaped electrodes in this reference are separated by 50 μm, so we speculate that this is maybe the reason why the $PC_L$ has not ever been noticed. Meanwhile, The location of the absorption edge ($PC_L$) also agrees well with the reference (DR spectrum, 1.48 eV)[10] and indicates that the photons with the energy around 1.48 eV could be absorbed by the materials and contribute to the generation of PC peak with an extremely low possibility of recombination. This is meaningful for realizing high-efficient photovoltaic devices based on such special materials.

The inset in **Figure 1c** shows the typical current-voltage (I-V) curves of the MSC in dark and under illumination with 3 mW cm$^{-2}$. It is apparent that both the dark current and photocurrent increase with the enhancement in voltage. The photocurrent value is three orders of magnitude higher than the dark current value and indicates a high efficiency of generation of e-h pairs by photon absorption. More detailed PC spectra under various electric fields with an incident light perpendicular to sample ($\theta = 0°$) at room temperature are given in **Figure 1c**. With electric field varying from 0 to 48 V/cm, the peak intensities of $PC_H$ and $PC_L$ increase almost linearly while corresponding positions are almost immobile (**Figure 1d**). All these linear relations indicate that the contact of the electrodes with MSC presents a typical Ohmic feature. At room temperature, no noticeable shift in the position of $PC_L$ was observed over a range of excitation powers (**Figure S5a**), and **Figure S5b** displays the dependence of $PC_L$ intensity on the excitation powers recorded. We observed a perfectly linear relationship between the power and the $PC_L$ intensity, which confirms that the absence of nonradiative recombination at room temperature and excitation power has almost no effect on the spectral shape of the PC spectrum.[23]

**Temperature-dependent PC-R and PL spectra.** To get more information on the unusual band structure, series of temperature-dependent spectroscopies was measured. **Figure 2** shows the according PC, R and PL spectra of the MSCs under temperatures ranging from 10 K to 300 K. Considering that $MAPbI_3$ presents a second-order phase transition from tetragonal to cubic phase at circa 330 K,[24, 25] we place the according spectra in **Figure S6** to make the figure display concisely. For the PC and R spectra, all the spectra show a prominent dependence of peak positions on the measuring temperatures, and the positions of two PC peaks are in agreement with those of the two absorption edges in R spectra. Meanwhile, the PL spectra of orthorhombic-$MAPbI_3$ possess three PL peaks, marked as $PL_H(O)$, $PL_L(O)$ and $PL_M(O)$ (**Figure 2e**). A new $PL_M(O)$ peak that appears at 1.6 eV between $PL_H(O)$ and $PL_L(O)$ also shows blueshift, which has been observed previously[22] when the temperature is below 45 K. In contrast to the orthorhombic phase, the tetragonal-$MAPbI_3$ possesses only one PL feature, marked as $PL_H(T)$ (**Figure 2f**).

To begin with the discussion, we have to determine the energy transition positions in these three sets of spectroscopic measurements. For R spectra, interestingly, we observe that high-energy part is sensitive to the angle ($\alpha$) between the main reflected light and the test direction of silicon detector while the low-energy part is inert to this angle (**Figure S7**). This indicates that the high-energy part is related to the specular reflection components and the low-energy part is relevant to the diffuse reflection components.[26,27] Unlike in PC and PL spectra where the transition energy could be determined conveniently using the emission peak position directly, the determination of the onset energy in R spectra becomes complicated, attributing to the two close absorption edges. Herein, we directly define the peak position of $R_H$ as the onset energy of high-energy absorption edge due to the obvious peak appearance, and use of $R_L$ to obtain the onset energy of low-energy absorption edge.[13] The dotted line indicates the linear extrapolation. As shown in **Figure S8**, a typical representation of these onset energies obtained from R spectrum at 78 K.

We then summarize the temperature dependences of PC and PL peak energies and R onset energies in **Figure 3**. The significant variations between 130 K and 150 K are attributed to structural phase transition of $MAPbI_3$ from low-temperature orthorhombic phase to the high-temperature tetragonal phase.[14,28] Apparent changes related to the second-order phase transition are not observable in **Figure 3**, but the obvious changes in peak intensity of $PC_L$ at around 300 K could be observed in **Figure S9**, indicative of an additional change from the tetragonal to cubic phase. The phase transitions of $MAPbI_3$ have been extensively studied, we would not focus on this issue.

In the temperature region lower than 130 K where MSC is in orthorhombic phase, two sets of transition energies，high-energy structure corresponding to $PC_H$, $R_H$ and $PL_H$ and low-energy structure corresponding to $PC_L$, $R_L$ and $PL_L$ coexist. Noticeably, $PL_L$ shows a feature of blueshift while $PC_L$ and $R_L$ present a redshift feature. This implicates that the corresponding band edge transition for $PL_L$ emission cannot be the same with that for $PC_L$ and $R_L$. The consistent blueshift tendency for $PL_H$, $PC_H$ and $R_H$ points to the same band edge for these transitions. In the temperature region where MSC is in tetragonal phase, though the onset energies change in comparison with those in orthorhombic phase, two sets of transition energies still coexist. Therefore, in high-energy part, the coincident blueshift trend of $PL_H$, $PC_H$ and $R_H$ indicates the same band edge for these transitions. In low-energy region, however, the $PL_L$ disappears while $PC_L$ and $R_L$ still show a redshift feature toward the increase in temperature.

The blue shift of $PC_H$, $PL_H$ and $R_H$ towards temperature increasing in both orthorhombic and tetragonal phases is opposite to the well-known Varshni relation, probably due to the relevant change in material structures and the interplay between the electron-phonon renormalization and thermal expansion.[28, 29] Since this observation has been well reported, we would emphasize on temperature evolution of $PC_L$, $PL_L$ and $R_L$. The $PC_L$ and $R_L$ are always in consistency and both show a redshift evolution with the increase of temperature, in agreement with the Varshni relation of normal inorganic semiconductors perfectly. However, the $PL_L$ corresponding to $PC_L$ and $R_L$ is not observable. The exciton binding energy of $MAPbI_3$ only has a small value of 30 meV at 13 K and the value decreases to 6 meV at 300 K.[30, 31, 32] The energy discrepancy between $PC_H$ and $PC_L$ ($\Delta PC$) varies from 50 meV (160 K) to 100 meV (300 K) and these values are prominently larger than those from excitonic effect measured at room temperature. Additionally, the strong signals of $PC_L$ and $R_L$ also indicate that such energy transition should not be originated from the defect or impurity levels. As shown in **Figure S10**, the full widths at half-maximum (FWHM) of $PL_H$ and $PC_L$ can be fitted by taking into account the temperature-independent inhomogeneous broadening ($\Gamma_0$) and the interaction between charge carriers and LO-phonons, described by the Fröhlich

Hamiltonian.[22, 23, 33] Besides, the PL spectra of MSCs might be influenced by the photonrecycling effect.[22, 34, 35, 36] Therefore, the temperature-dependent $\Gamma_{PC}$ reflects more accurately than the temperature-dependent $\Gamma_{PL}$. The extracted fitting values of these peaks are shown in **Table S1**, which indicate that not the low-energy structure (PC$_L$) but the high-energy structure (PL$_H$) agrees with the literatures.[22, 23, 33] All these point out that in MAPbI$_3$ there exists a special band structure where the photons can be absorbed and converted to charge carriers efficiently with an extremely low possibility for recombining radioactively. This is crucial to the materials for acquiring a high-efficient photovoltaic effect.

**The kinetic model of the absorption and photoluminescence processes.** Herein, the coexistence of two sets of transition energies in both crystal phases indicates that this is not governed by the crystal structure and a more fundamental issue like the ionic crystal nature could be responsible for this observation. Thus, the conventional consensus, that MA group does not have any significant contribution to the electronic structure around the band edges and the only role is donating electrons to the Pb-I framework, is not feasible and somewhat inaccurate in calculation.[37, 38, 39] More theoretical investigations suggest that MA molecule rotations, with the consequent dynamical change in the band structure, might be the origin of the slow carrier recombination and the superior power conversion efficiency of MAPbI$_3$ based photovoltaic devices.[15, 40, 41, 42] Dar *et al.* recently claimed that the two PL peaks are associated with the MA-ordered and MA-disordered domains in MAPbI$_3$, via the density functional theory calculations in combination with molecule dynamics models.[23]

In fact, MA cations play an important role in these anomalous emission characteristics of the lead halide perovskites. For calculating the energy band structures, all the researchers assumed that the PbI$_6$ octahedral is a rigid body, MA only affects the tilting of the iodine octahedron in these theoretical calculations, and the bond length and the bond angle of Pb-I-Pb in the octahedron do not change[15, 24, 40, 41, 42, 43]. However, due to the electrostatic interactions of MA and the surrounded P-I bond, in reality, the bond length and bond angle of Pb-I-Pb in the octahedron should have some adjustments according to the acentric electrostatic nature of MA cation. Therefore, it is of necessity to consider the change of P-I bond in the different positions of the surrounding MA cation.

Specifically in hybrid perovskites, hydrogen bonding (N-H⋯I) effect between the amine group and the halide ions has been verified both theoretically and experimentally.[24, 44, 45, 46] In this regard, MA molecule would be immobilized by hydrogen bonds and cannot rotate randomly, herein we call it MA-lock state. The distances between the amine group and the surrounding halide ions should not be the same, and the Pb-I bond length near the amine group becomes shorter due to hydrogen bonding and the Pb-I bond length away from the amine group becomes longer under the MA-lock state. As schematically shown in **Figure 4a**, the hydrogen bonding between N and I is geometrically coupled to the buckling of the Pb-I-Pb bond and the changing of the Pb-I bond length. Under the low-energy optical absorption, the first direct transition corresponds to the charge transfer from hybridized Pb *6s*-I *5p* orbital to the Pb *6p* orbital, forming weakly bound excitons. [47, 48, 49] Accordingly, the electronic transition from I atom to Pb atom directly leads to the reduction of electron density on the I site, and thereby reduces its Coulomb interaction with the amine group. This in turn straightens the Pb-I-Pb bond due to the lattice relaxes, resulting in the MA-unlock state in which the MA molecule turns into the condition of random rotation (**Figure 4b**). Comparing to the long lifetime of charge carriers (22~1032 ns),[8] the vibrational (~300 fs) and rotational (~3 ps) motion of the organic cation[50] are very fast. In this case, the electrostatic forces of MA to all the surrounding Pb-I bonds should be the same, thus the length and angle of all the Pb-I bonds should be equal. Furthermore, MA-unlock state would also immediately turn into MA-lock state once again after carrier recombination or extraction.

Based on the proposed molecular dynamic model, there are two types of Pb-I frameworks under MA-lock state which may correspond to high-energy and low-energy band structures. The corresponding two energy band structures are schematically depicted in **Figure 4c**. Here, there are two direct band structures with the two conduction bands ($CB_H$ and $CB_L$) corresponding to the two valence bands ($VB_H$ and $VB_L$) in $k$-space, resulting in two absorption peaks. However, under MA-unlock state, only one type of Pb-I framework exists and contributes to one band structure (**Figure 4d**). In MA-unlock state, the carrier recombination has only one luminescence peak after optical absorption. This nature of the changeable band structure in MSC is responsible to the fact that $PL_L$ is even not observable.

The feasibility of this model could be verified by the following observations. First, in the presence of changeable band structure, the photo-generated electrons and holes in high-energy band structure have a higher possibility to recombine than the electrons and holes in low-energy band structure due to the disappearance of the low-energy band structure in MA-unlock state. This results in a high $PC_L$ signal rather than $PC_H$. Second, this model also indicates carrier could relax from the band structure in **Figure 4c** to the band structure in **Figure 4d** by absorbing phonon energy. This process reduces all possible carrier recombination rates, which thus contributes to the experimentally observed long carrier lifetime and diffusion length in the hybrid lead halide perovskites.[11] Finally, the kinetics on photo-involved charge transition and transfer are investigated using the pump-probe photoconductivity technique.

**The Pump-probe PC spectroscopy.** To check the feasibility of this model, we designed a system of Pump-probe PC spectroscopy to monitor the dynamics of charge generation and separation, as shown in inset of **Figure 5**. Meanwhile, **Figure 5** shows the statistical results of the intensity change in $PC_L$ peak ($\Delta PC_L$, probe light), corresponding to the pump photon energy. The two curves measured at 78 K and 293 K (more temperature-dependent Pump-probe PC spectra are shown in **Figure S11**), respectively, present a similar contour. Specifically, as the pump light energy is the same as $PC_L$ peak energy, the attenuation of $PC_L$ shows the highest level. On the contrary, as the pump light energy is close to $PC_H$ peak energy, $PC_L$ signal shows an increasing feature. In the spectroscopic region of $PC_L$ attenuation, the occupancies for the large number of electron-hole pairs generated by the pump light on the low-energy band edge would push the electrons and holes generated by the probe light to the high-energy band edge (**Figure 4c**) and thus deteriorate the PC signal around $PC_L$ peak. In the region of $PC_L$ enhancement, the large amount of electrons and holes excited by pump light is of high possibility to occupy the vacancy on the high-energy band edge and this would inhibit the transfer of charges generated by the probe light to the high-energy band edge (**Figure 4c**). In this case, the recombination of e-h (probe light) would be inhibited, resulting in the increasing feature of $PC_L$. Therefore, the feasibility of our model on the two energy transition channels has been verified. The existence of another band edge below the "true" band edge reported by many research groups [11, 12] has been given the definite physical meaning and causes the complicated optical responses. The charge transitions from the exited low-energy channel to the high-energy channel by the assistance of additional photon or phonon could also result in interesting effects of photoluminescence upconversion or laser cooling. The according phenomena have been well reported[51] and our model provides a direct physical support for the observations.

# ■ CONCLUSIONS

In conclusion, we have studied the optical properties of MSCs by means of PC, R and PL spectra. Although the unusual blue shifts of $PC_H$, $R_H$ and $PL_H$ have been observed previously, this low-energy $PC_L$ and $R_L$ peaks and their red shifts are the first discoveries on MSCs. More importantly, the according band structure shows an ignorable possibility of charge recombination. It is envisaged that solar cells made of single-crystalline perovskites will render significantly higher power conversion efficiency, as it offers not only better carrier generation and transport efficiencies, but also a broader light absorption spectrum. This work is essential for further development of highly efficient solar cells and other optoelectronic devices based on organometal halide perovskites.

■ **ACKNOWLEDGMENT**

This work was supported by the National Natural Science Foundation of China (Contract No. 11574302, 61474114, 61674141, 61504134, 21503209 and 11704032), the National Basic Research Program of China (Grant No. 2015CB921503, 2014CB643503 and 2013CB632805), National Key Research and Development Program (Grant No. 2016YFB0402303 and 2016YFB0400101), Key Research Program of Frontier Science, CAS (Grant No. QYZDB-SSW-SLH006).

■ **AUTHOR INFORMATION**

**Author contributions**

W.H. and S.Y. contributed equally to this work. W.H. and Y.C. conceived the experiment and analysed the data. W.H. performed the experiments. S.Y. S.Q. and Z.W. provided $CH_3NH_3PbI_3$ perovskite single crystals. S.Y., P.L. and W.H. performed device fabrication. W.H., Z.W. and Y. L. wrote the manuscript. All the authors discussed the results and commented on the manuscript.

**Corresponding Author**

E-mail: yhchen@semi.ac.cn.

E-mail: wangzj@semi.ac.cn.

**Notes**

The authors declare no competing financial interests.

■ **REFERENCES**

1. Yang WS, Noh JH, Jeon NJ, Kim YC, Ryu S, Seo J, *et al.* High-performance photovoltaic perovskite layers fabricated through intramolecular exchange. *Science*. **348**, 1234-1237 (2015).
2. Zhu H, Fu Y, Meng F, Wu X, Gong Z, Ding Q, *et al.* Lead halide perovskite nanowire lasers with low lasing thresholds and high quality factors. *Nature Materials*. **14**, 636-U115 (2015).
3. Lin Q, Armin A, Burn PL, Meredith P. Filterless narrowband visible photodetectors. *Nature Photonics*. **9**, 687 (2015).
4. Saidaminov MI, Adinolfi V, Comin R, Abdelhady AL, Peng W, Dursun I, *et al.* Planar-integrated single-crystalline perovskite photodetectors. *Nature Communications*. **6**, (2015).
5. Wetzelaer G-JAH, Scheepers M, Miquel Sempere A, Momblona C, Avila J, Bolink HJ. Trap-Assisted Non-Radiative Recombination in Organic-Inorganic Perovskite Solar Cells. *Advanced Materials*. **27**, 1837 (2015).
6. Mei A, Li X, Liu L, Ku Z, Liu T, Rong Y, *et al.* A hole-conductor-free, fully printable mesoscopic perovskite solar cell with high stability. *Science*. **345**, 295-298 (2014).
7. Shao Y, Xiao Z, Bi C, Yuan Y, Huang J. Origin and elimination of photocurrent hysteresis by fullerene passivation in CH3NH3PbI3 planar heterojunction solar cells. *Nature Communications*. **5**, (2014).
8. Shi D, Adinolfi V, Comin R, Yuan M, Alarousu E, Buin A, *et al.* Low trap-state density and long carrier diffusion in organolead trihalide perovskite single crystals. *Science*. **347**, 519-522 (2015).
9. Saidaminov MI, Abdelhady AL, Murali B, Alarousu E, Burlakov VM, Peng W, *et al.* High-quality bulk hybrid perovskite single crystals within minutes by inverse temperature crystallization. *Nature Communications*. **6**, (2015).
10. Dang Y, Liu Y, Sun Y, Yuan D, Liu X, Lu W, *et al.* Bulk crystal growth of hybrid perovskite material CH3NH3PbI3. *Crystengcomm*. **17**, 665-670 (2015).
11. Dong Q, Fang Y, Shao Y, Mulligan P, Qiu J, Cao L, *et al.* Electron-hole diffusion lengths > 175 μm in solution-grown CH3NH3PbI3 single crystals. *Science*. **347**, 967-970 (2015).
12. Fang H-H, Raissa R, Abdu-Aguye M, Adjokatse S, Blake GR, Even J, *et al.* Photophysics of Organic-Inorganic Hybrid Lead Iodide Perovskite Single Crystals. *Advanced Functional Materials*. **25**, 2378-2385 (2015).
13. Yasuhiro Y, Toru N, Masaru E, Atsushi W, Yoshihiko K. Near-band-edge optical responses of solution-processed organic–inorganic hybrid perovskite CH3NH3PbI3 on mesoporous TiO2 electrodes. *Applied Physics Express*. **7**, 032302 (2014).
14. Milot RL, Eperon GE, Snaith HJ, Johnston MB, Herz LM. Temperature-Dependent Charge-Carrier Dynamics in CH3NH3PbI3 Perovskite Thin Films. *Advanced Functional Materials*. **25**, 6218-6227 (2015).
15. Motta C, El-Mellouhi F, Kais S, Tabet N, Alharbi F, Sanvito S. Revealing the role of organic cations in hybrid halide perovskite CH3NH3PbI3. *Nature Communications*. **6**, (2015).
16. Zheng F, Tan LZ, Liu S, Rappe AM. Rashba Spin-Orbit Coupling Enhanced Carrier Lifetime in CH(3)NH(3)PbI(3). *Nano Letters*. **15**, 7794-7800 (2015).
17. Etienne T, Mosconi E, De Angelis F. Dynamical Origin of the Rashba Effect in Organohalide Lead Perovskites: A Key to Suppressed Carrier Recombination in Perovskite Solar Cells? *Journal of Physical Chemistry Letters*. **7**, 1638-1645 (2016).
18. Hutter EM, Gelvez-Rueda MC, Osherov A, Bulovic V, Grozema FC, Stranks SD, *et al.* Direct-

indirect character of the bandgap in methylammonium lead iodide perovskite. *Nature Materials*. **16**, 115-120 (2017).

19. Cho N, Li F, Turedi B, Sinatra L, Sarmah SP, Parida MR, *et al.* Pure crystal orientation and anisotropic charge transport in large-area hybrid perovskite films. *Nature Communications*. **7**, (2016).

20. Baikie T, Fang Y, Kadro JM, Schreyer M, Wei F, Mhaisalkar SG, *et al.* Synthesis and crystal chemistry of the hybrid perovskite (CH3NH3) PbI3 for solid-state sensitised solar cell applications. *Journal of Materials Chemistry A*. **1**, 5628-5641 (2013).

21. Liu Y, Yang Z, Cui D, Ren X, Sun J, Liu X, *et al.* Two-Inch-Sized Perovskite CH3NH3PbX3 (X = Cl, Br, I) Crystals: Growth and Characterization. *Advanced Materials*. **27**, 5176-5183 (2015).

22. Phuong LQ, Nakaike Y, Wakamiya A, Kanemitsu Y. Free Excitons and Exciton-Phonon Coupling in CH3NH3PbI3 Single Crystals Revealed by Photocurrent and Photoluminescence Measurements at Low Temperatures. *Journal of Physical Chemistry Letters*. **7**, 4905-4910 (2016).

23. Dar MI, Jacopin G, Meloni S, Mattoni A, Arora N, Boziki A, *et al.* Origin of unusual bandgap shift and dual emission in organic-inorganic lead halide perovskites. *Science advances*. **2**, e1601156-e1601156 (2016).

24. Zhou Y, You L, Wang S, Ku Z, Fan H, Schmidt D, *et al.* Giant photostriction in organic-inorganic lead halide perovskites. *Nature Communications*. **7**, (2016).

25. Poglitsch A, Weber D. Dynamic disorder in methylammoniumtrihalogenoplumbates (II) observed by millimeter-wave spectroscopy. *Journal of Chemical Physics*. **87**, 6373-6378 (1987).

26. Sharada G, Mahale P, Kore BP, Mukherjee S, Pavan MS, De CD, *et al.* Is CH3NH3PbI3 Polar? *Journal of Physical Chemistry Letters*. **7**, 2412-2419 (2016).

27. Kim H-S, Lee C-R, Im J-H, Lee K-B, Moehl T, Marchioro A, *et al.* Lead Iodide Perovskite Sensitized All-Solid-State Submicron Thin Film Mesoscopic Solar Cell with Efficiency Exceeding 9%. *Scientific Reports*. **2**, (2012).

28. Kong W, Ye Z, Qi Z, Zhang B, Wang M, Rahimi-Iman A, *et al.* Characterization of an abnormal photoluminescence behavior upon crystal-phase transition of perovskite CH3NH3PbI3. *Physical Chemistry Chemical Physics*. **17**, 16405-16411 (2015).

29. Shi ZF, Sun XG, Wu D, Xu TT, Tian YT, Zhang YT, *et al.* Near-infrared random lasing realized in a perovskite CH3NH3PbI3 thin film. *Journal of Materials Chemistry C*. **4**, 8373-8379 (2016).

30. Yamada Y, Nakamura T, Endo M, Wakamiya A, Kanemitsu Y. Photoelectronic Responses in Solution-Processed Perovskite CH3NH3PbI3 Solar Cells Studied by Photoluminescence and Photoabsorption Spectroscopy. *Ieee Journal of Photovoltaics*. **5**, 401-405 (2015).

31. Yamada Y, Nakamura T, Endo M, Wakamiya A, Kanemitsu Y. Photocarrier Recombination Dynamics in Perovskite CH3NH3PbI3 for Solar Cell Applications. *Journal of the American Chemical Society*. **136**, 11610-11613 (2014).

32. Ziffer ME, Mohammed JC, Ginger DS. Electroabsorption Spectroscopy Measurements of the Exciton Binding Energy, Electron–Hole Reduced Effective Mass, and Band Gap in the Perovskite CH3NH3PbI3. *ACS Photonics*. **3**, 1060-1068 (2016).

33. Wright AD, Verdi C, Milot RL, Eperon GE, Perez-Osorio MA, Snaith HJ, *et al.* Electron-phonon coupling in hybrid lead halide perovskites. *Nature Communications*. **7**, (2016).

34. Yamada Y, Yamada T, Le Quang P, Maruyama N, Nishimura H, Wakamiya A, *et al.* Dynamic Optical Properties of CH(3)NH(3)PbI(3) Single Crystals As Revealed by One- and Two-Photon Excited Photoluminescence Measurements. *Journal of the American Chemical Society*. **137**, 10456-

10459 (2015).

35. Yamada T, Yamada Y, Nakaike Y, Wakamiya A, Kanemitsu Y. Photon Emission and Reabsorption Processes in CH3NH3PbBr3 Single Crystals Revealed by Time-Resolved Two-Photon-Excitation Photoluminescence Microscopy. *Physical Review Applied*. **7**, (2017).

36. Yamada T, Yamada Y, Nishimura H, Nakaike Y, Wakamiya A, Murata Y, *et al.* Fast Free-Carrier Diffusion in CH3NH3PbBr3 Single Crystals Revealed by Time-Resolved One- and Two-Photon Excitation Photoluminescence Spectroscopy. *Advanced Electronic Materials*. **2**, (2016).

37. Wang L, Wang K, Xiao G, Zeng Q, Zou B. Pressure-Induced Structural Evolution and Band Gap Shifts of Organometal Halide Perovskite-Based Methylammonium Lead Chloride. *The Journal of Physical Chemistry Letters*. 5273-5279 (2016).

38. Menendez-Proupin E, Palacios P, Wahnon P, Conesa JC. Self-consistent relativistic band structure of the CH3NH3PbI3 perovskite. *Physical Review B*. **90**, (2014).

39. Umari P, Mosconi E, De Angelis F. Relativistic GW calculations on CH3NH3PbI3 and CH3NH3SnI3 Perovskites for Solar Cell Applications. *Scientific Reports*. **4**, (2014).

40. Yun S, Zhou X, Even J, Hagfeldt A. Recent progress of first principles calculations in CH3NH3PbI3 perovskite solar cells. *Angewandte Chemie International Edition*. n/a (2017).

41. Goehry C, Nemnes GA, Manolescu A. Collective Behavior of Molecular Dipoles in CH3NH3PbI3. *Journal of Physical Chemistry C*. **119**, 19674-19680 (2015).

42. Chen HA, Lee MH, Chen CW. Wavelength-dependent optical transition mechanisms for light-harvesting of perovskite MAPbI(3) solar cells using first-principles calculations. *Journal of Materials Chemistry C*. **4**, 5248-5254 (2016).

43. Berdiyorov GR, Kachmar A, El-Mellouhi F, Carignano MA, Madjet ME. Role of Cations on the Electronic Transport and Optical Properties of Lead-Iodide Perovskites. *Journal of Physical Chemistry C*. **120**, 16259-16270 (2016).

44. Lee J-H, Bristowe NC, Bristowe PD, Cheetham AK. Role of hydrogen-bonding and its interplay with octahedral tilting in CH3NH3PbI3. *Chemical Communications*. **51**, 6434-6437 (2015).

45. Swainson I, Chi L, Her J-H, Cranswick L, Stephens P, Winkler B, *et al.* Orientational ordering, tilting and lone-pair activity in the perovskite methylammonium tin bromide, CH3NH3SnBr3. *Acta Crystallographica Section B-Structural Science*. **66**, 422-429 (2010).

46. Glaser T, Mueller C, Sendner M, Krekeler C, Semonin OE, Hull TD, *et al.* Infrared Spectroscopic Study of Vibrational Modes in Methylammonium Lead Halide Perovskites. *Journal of Physical Chemistry Letters*. **6**, 2913-2918 (2015).

47. Miyata A, Mitioglu A, Plochocka P, Portugall O, Wang JT-W, Stranks SD, *et al.* Direct measurement of the exciton binding energy and effective masses for charge carriers in organic-inorganic tri-halide perovskites. *Nat Phys*. **11**, 582-587 (2015).

48. D'Innocenzo V, Grancini G, Alcocer MJP, Kandada ARS, Stranks SD, Lee MM, *et al.* Excitons versus free charges in organo-lead tri-halide perovskites. *Nature Communications*. **5**, (2014).

49. Saba M, Cadelano M, Marongiu D, Chen F, Sarritzu V, Sestu N, *et al.* Correlated electron-hole plasma in organometal perovskites. *Nature Communications*. **5**, (2014).

50. Bakulin AA, Selig O, Bakker HJ, Rezus YLA, Mueller C, Glaser T, *et al.* Real-Time Observation of Organic Cation Reorientation in Methylammonium Lead Iodide Perovskites. *Journal of Physical Chemistry Letters*. **6**, 3663-3669 (2015).

51. Ha S-T, Shen C, Zhang J, Xiong Q. Laser cooling of organic-inorganic lead halide perovskites. *Nature Photonics*. **10**, 115 (2016).

# ■ FIGURES

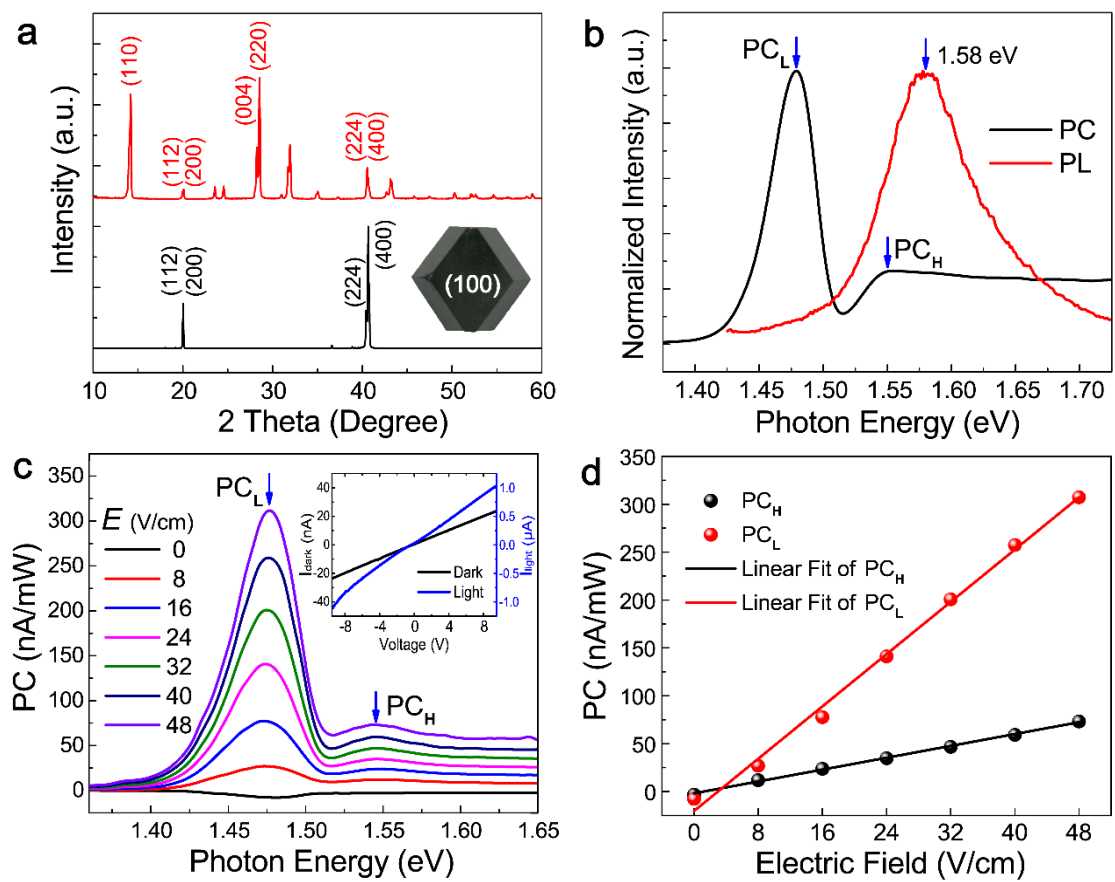

**Figure 1 | Basic properties of the MSCs at room temperature (293 K). a,** XRD patterns of the large perovskite MSC (black) and the ground powder from large MSC (red) with the photograph of MSC shown in the down-right inset. **b,** Normalized PC (the electric field is 32 V/cm) and PL spectra of the MSC with the incident light perpendicular to sample. **c,** Electric field-dependent PC spectra with an incident light perpendicular to sample with the I-V curves of MSC shown in the upper-right inset. **d,** The statistical results of the peak intensity of $PC_H$ and $PC_L$ from figure 1c.

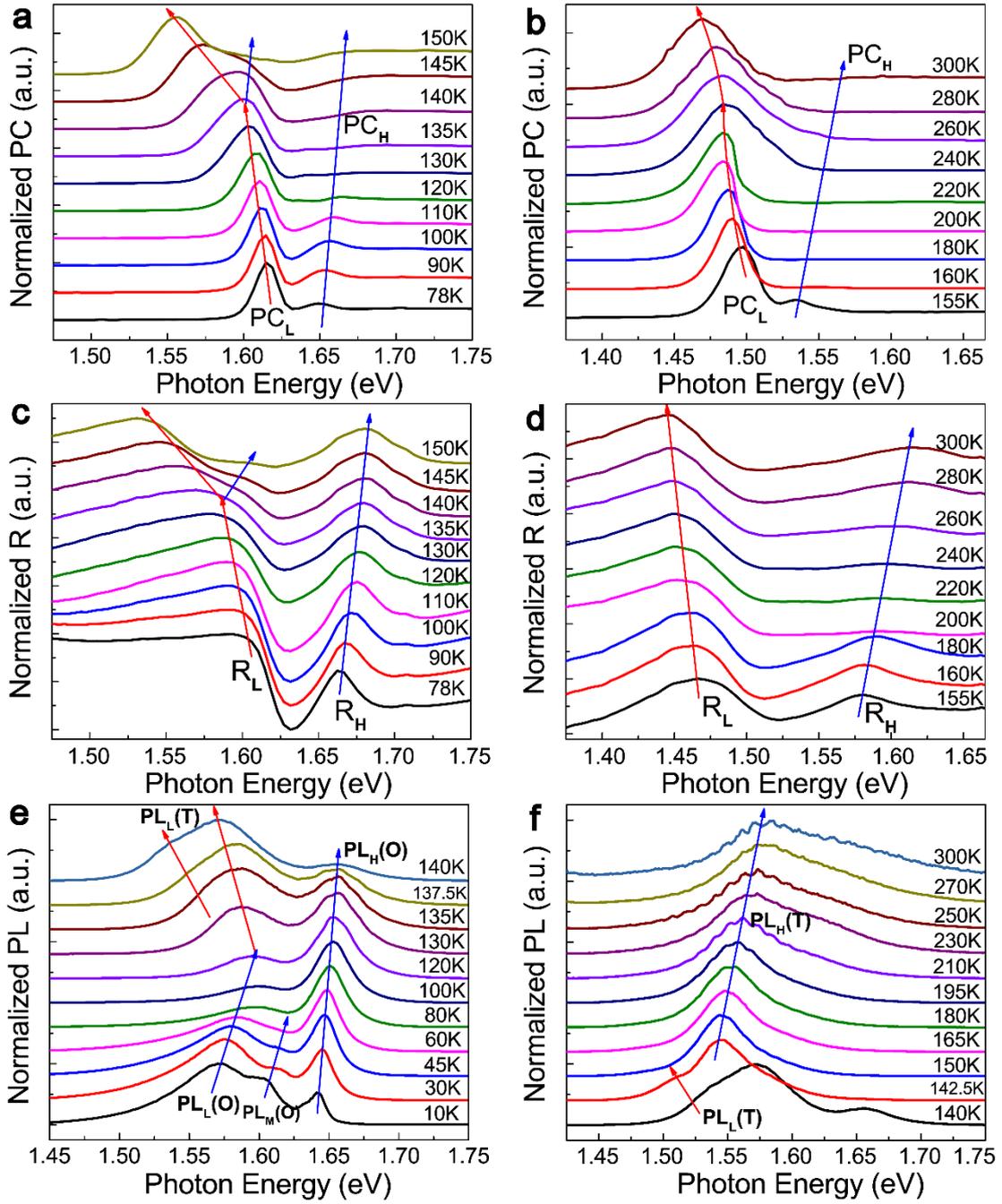

**Figure 2 | Temperature-dependent spectra of the MSC. a,b,** The normalized PC spectra at different temperatures with an external electric field of 32 V/cm. **c,d,** Normalized R spectra with the incident light oblique incidence to sample ($\theta = 10°$). **e,f,** Normalized PL spectra from 10 K to 300 K. The red (blue) arrow lines in the figure indicate that the peaks exhibit continuous redshifts (blueshifts) as temperature increases. All the spectra are intentionally shifted for ease of viewing.

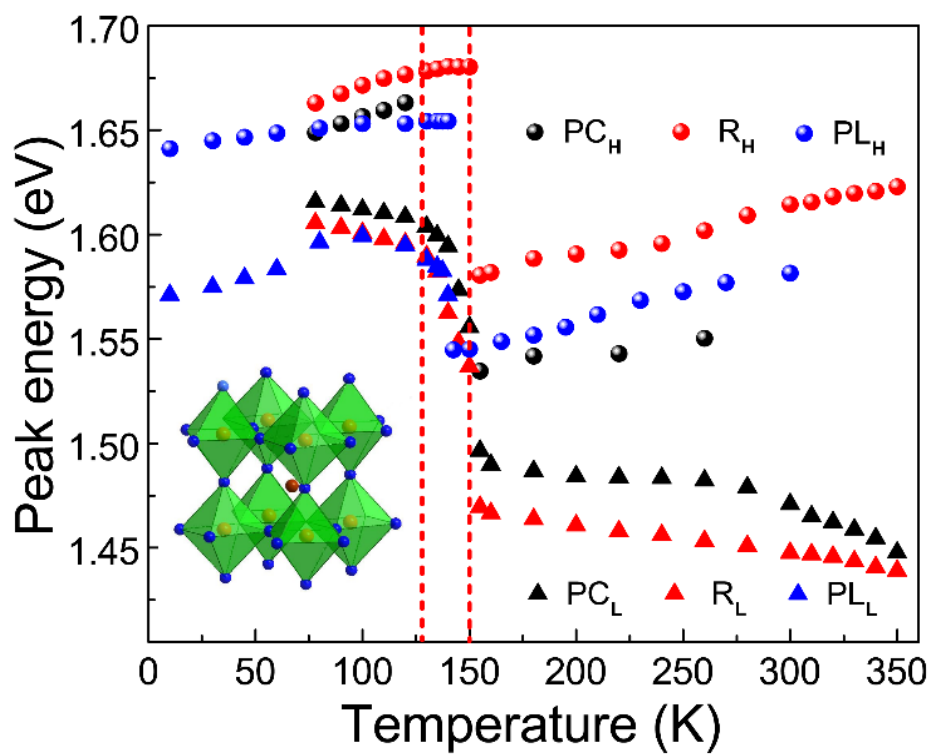

**Figure 3 | The band edge characteristics of MSC.** The temperature dependences of PC and PL peak energies and R onset energies. The left inset shows the crystal structure of tetragonal-MAPbI$_3$ phase. The rigid framework consisting of the PbI$_6$ octahedral is also highlighted in the structure.

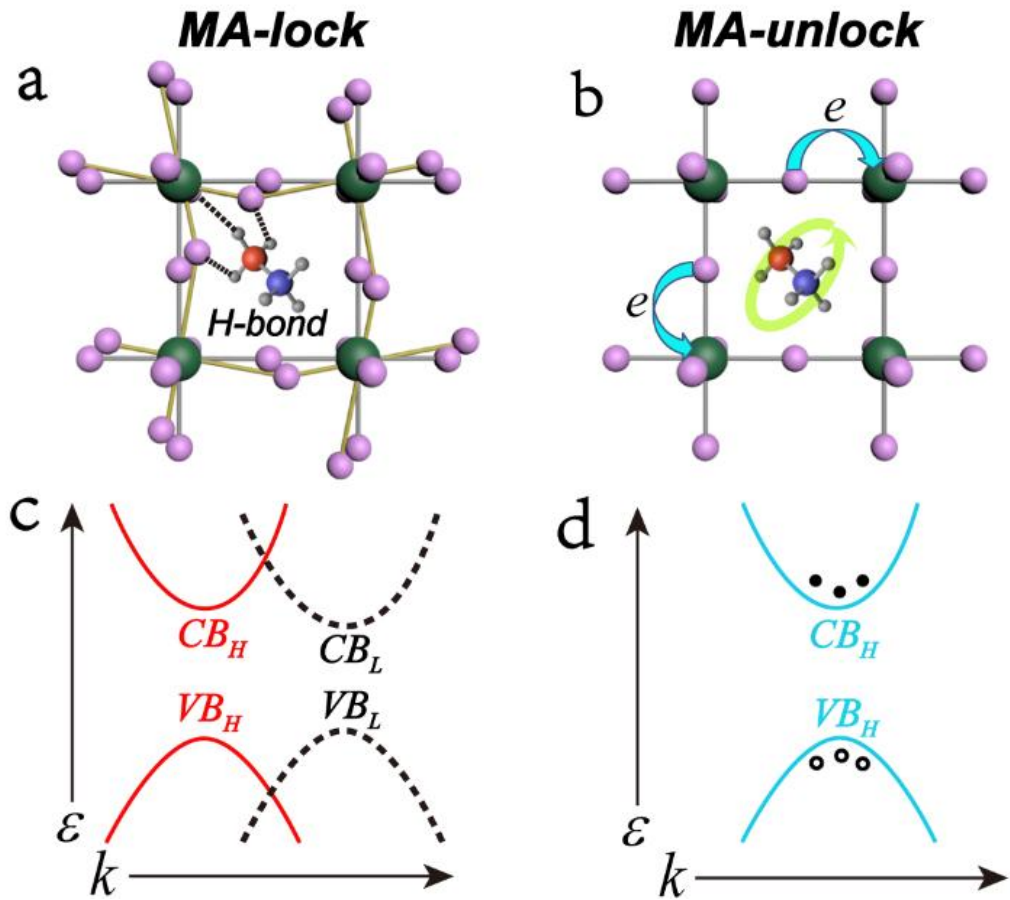

**Figure 4 | Proposed mechanism for the absorption and photoluminescence processes in MSC.** (a-b) Molecular dynamics schematic illustrations (not to scale) showing that the MA-lock state under the hydrogen bonding between the amine group and the iodine ion and the MA-unlock state that the MA molecule turn into the random rotation, respectively. (c-d) Proposed band diagram correspond to the MA-lock state and the MA-unlock state for MAPbI$_3$ single crystal, respectively.

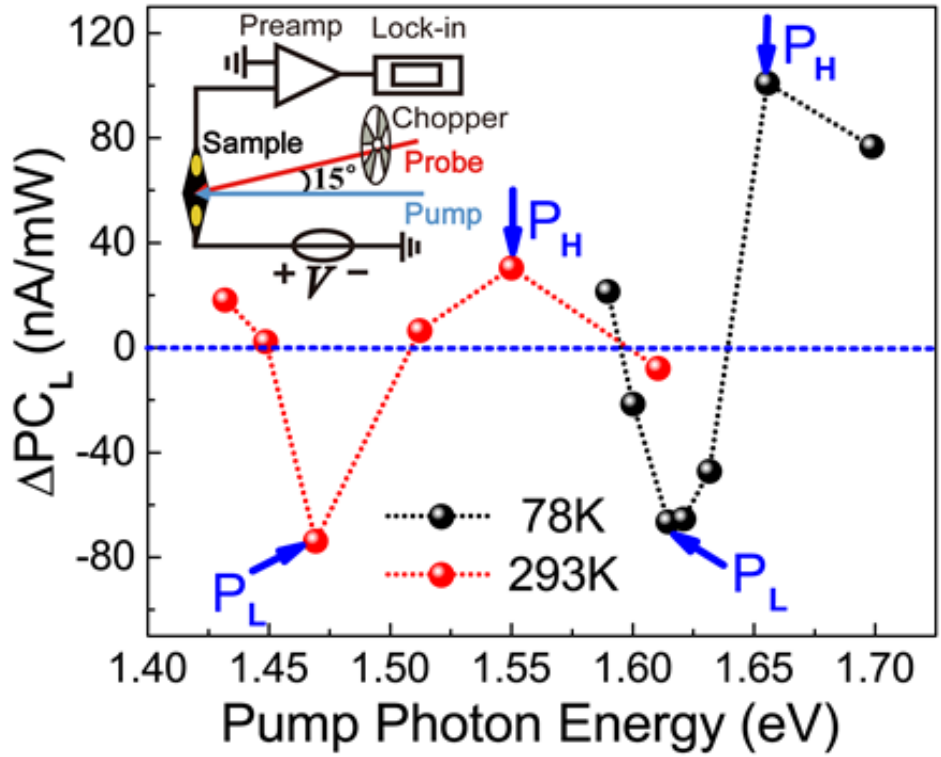

**Figure 5 | The Pump-probe PC spectroscopy.** Pump light photon energy dependence of the change in $PC_L$ intensity ($\Delta PC_L$, probe light) at 78 K (black) and 293 K (red). The upper-left inset shows the schematic diagram of Pump-probe PC spectroscopy.

■ **SUPPORTING INFORMATION**

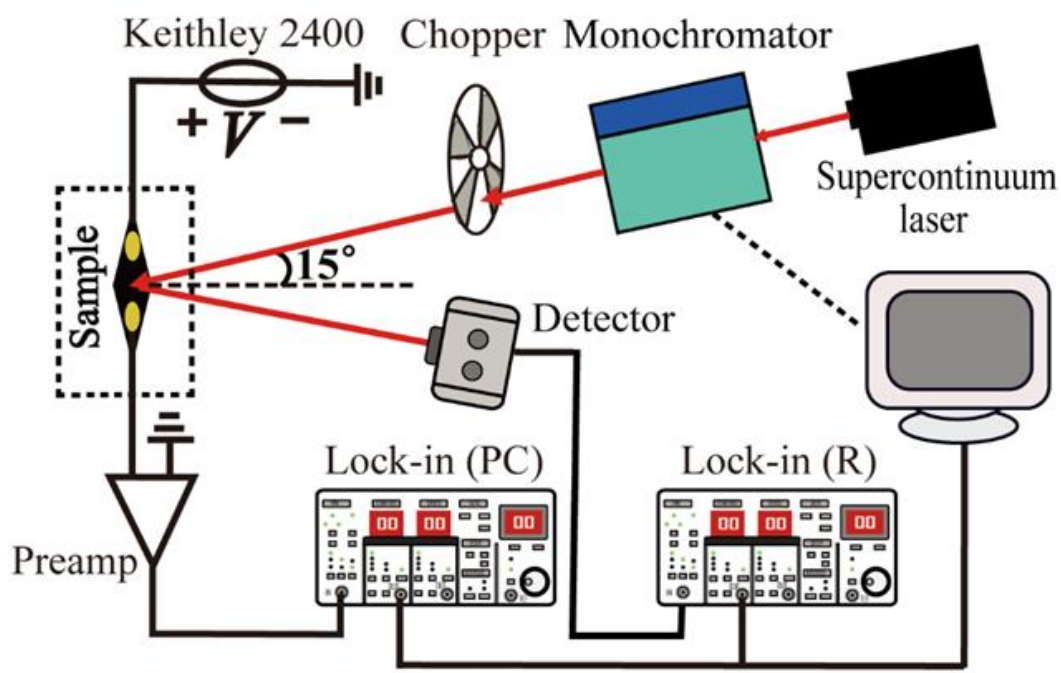

**Figure S1.** The schematic of our PC-R system setup.

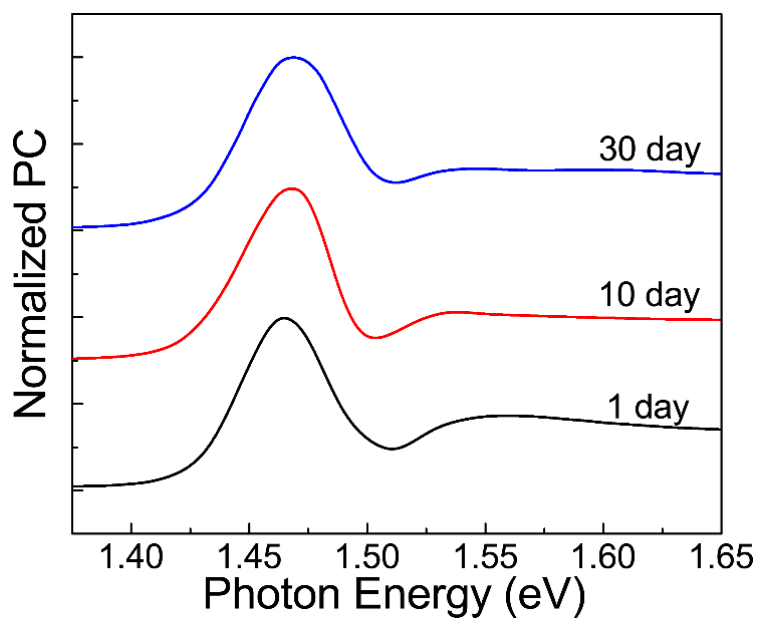

**Figure S2.** The normalized PC spectra of MSC with different storage time at room temperature. The sample is mounted into an optical cryostat under vacuum ($10^{-5}$ Torr), and all the spectra are intentionally shifted for ease of viewing.

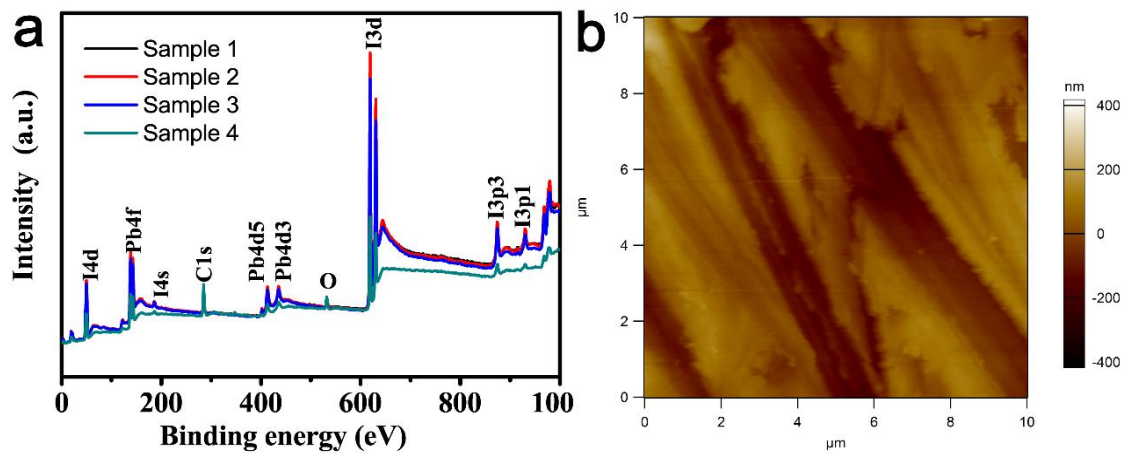

**Figure S3.** The surface properties of MSC. (a) XPS of four MSC samples. The surface of Sample 1 has not undergone any treatment, the Sample 2 is thinned by 0.1 mm and polished along one direction, the Sample 3 is thinned by 1 mm and polished along one direction, and the Sample 4 is fabricated by depositing Au and wiring bonding from Sample 3. (b) The AFM image of the sample 3, and the average roughness of this area is about 70 nm.

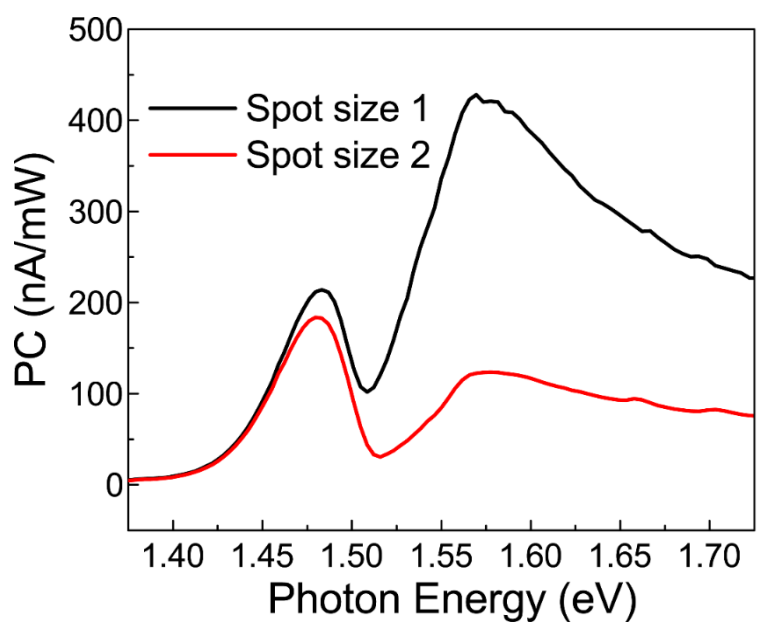

**Figure S4.** The PC spectra of MSC under different light spot sizes. The size of light spot 2 is less than the electrode spacing (red curve), and a small amount of light spot 1 illuminates the electrodes (black curve).

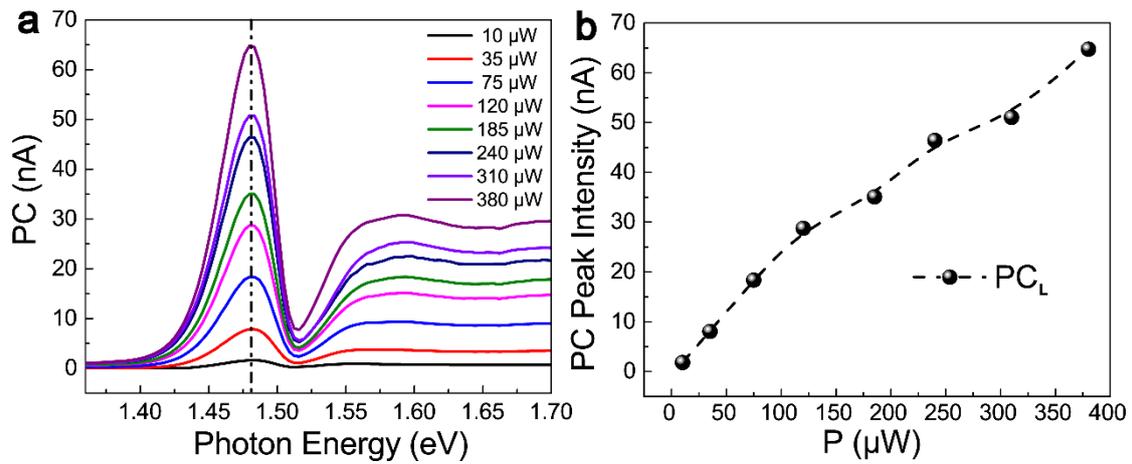

**Figure S5.** Excitation power-dependent PC spectra at room temperature. (a) The PC spectra of MSC under different excitation powers. (b) The statistical results of the peak intensity of $PC_L$ from (a).

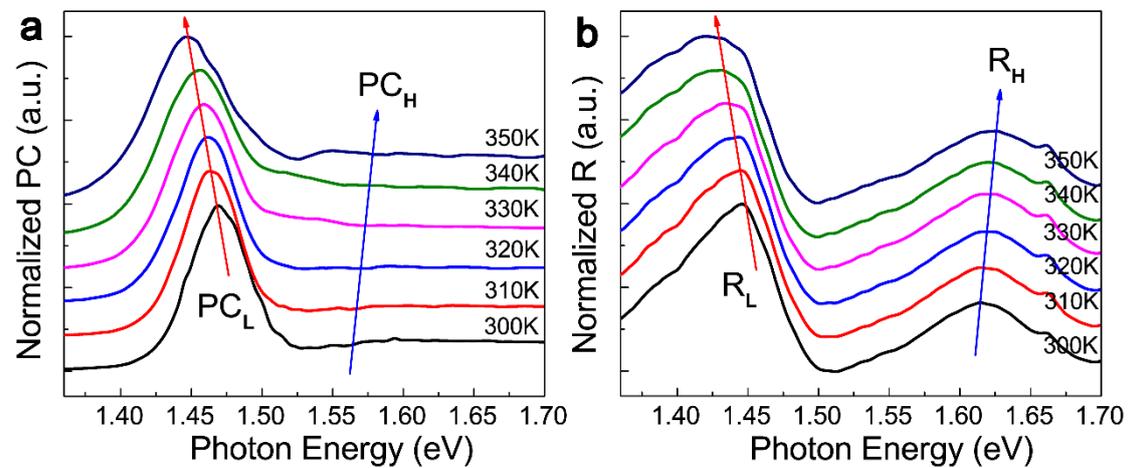

**Figure S6.** Temperature-dependent spectra of the MSC measured from 300 to 350 K. (a) The normalized PC spectra at different temperatures with an external electric field of 32 V/cm. (b) The normalized R spectra of the same sample with an oblique light incidence ($\theta = 10°$).

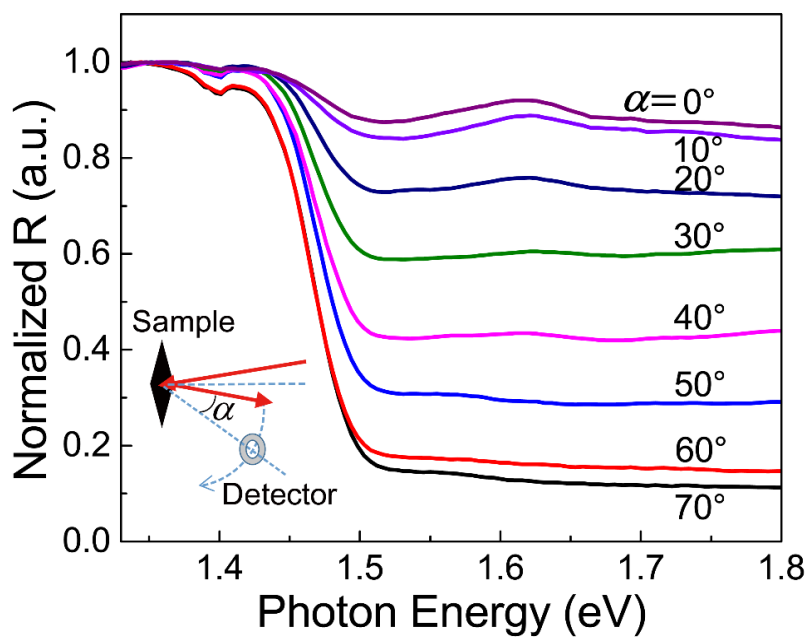

**Figure S7.** *α* angle dependence of the normalized R (or DR) spectra of the MSC with the schematic illustration of the geometry of the validation experiment shown in the down-left inset, where *α* is the angle between the main reflected light and the test direction of Si detector.

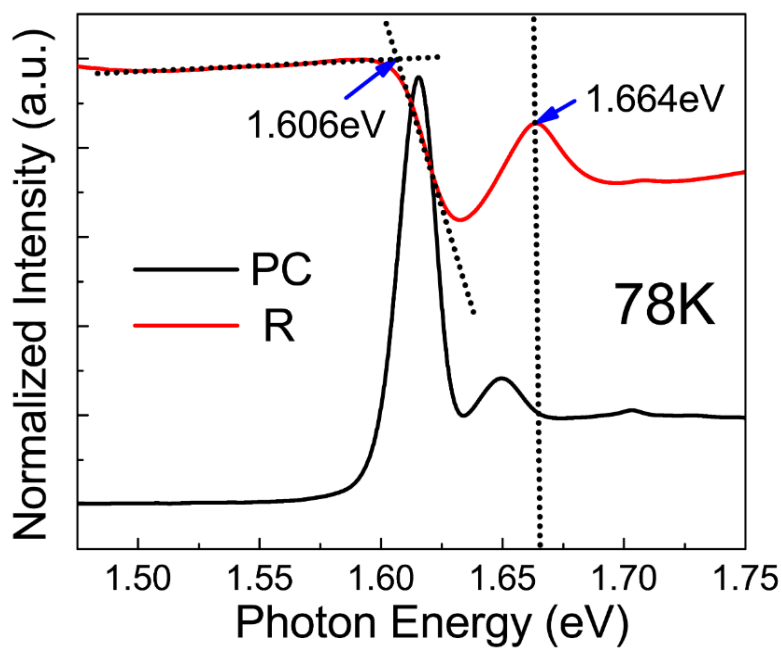

**Figure S8.** A typical representative quantitative estimation of the absorption edge through PC and R spectra of MSC at 78 K.

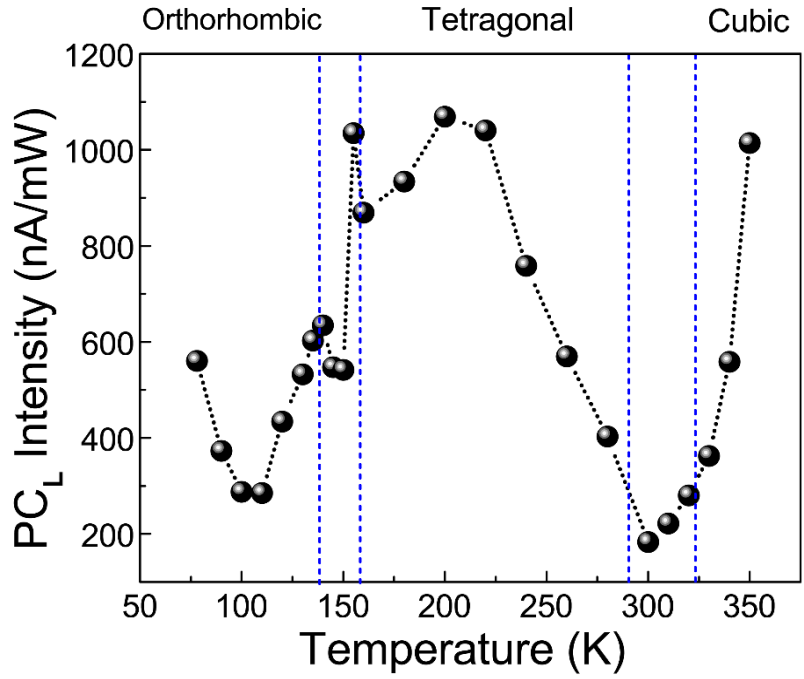

**Figure S9.** The temperature dependence of the intensity of $PC_L$.

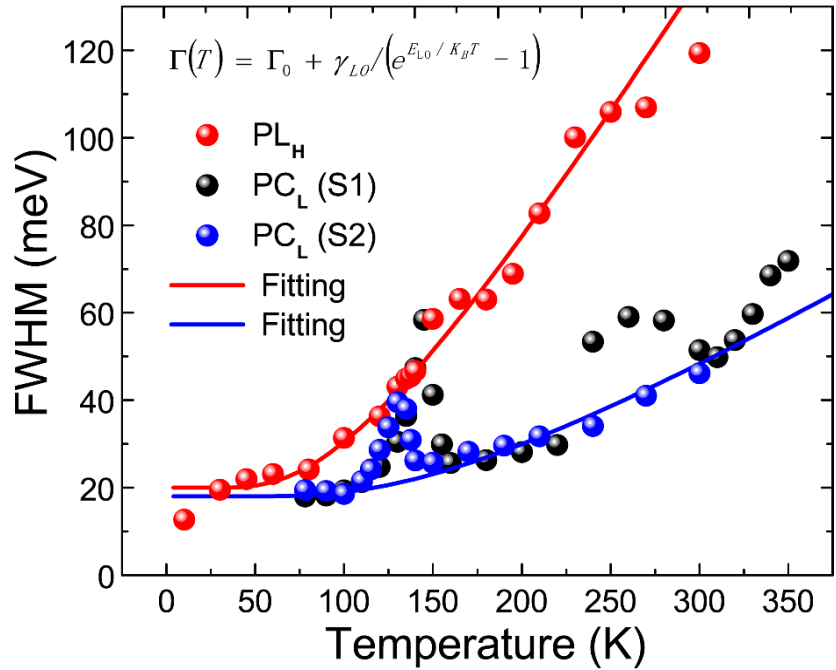

**Figure S10.** FWHMs of the $PL_H$ peak (red circles) and the $PC_L$ peak (black and blue circles) obtained for $MAPbI_3$ single crystals as functions of temperature. The solid red and blue lines show the fit for the temperature-dependent $\Gamma_{PL}$ and $\Gamma_{PC}$.

Firstly, we obtained the PC spectrum (see Figure S11a) of the sample without pump light at 78 K (Orthorhombic phase). In this case, we could determine the characteristic energies corresponding to the high-energy structure and low-energy structure of the sample, respectively. Finally, we measured Pump-probe PC spectra (see Figure S11b) under pump light modulation at different photon energies. Similarly, Pump-probe PC spectra (see Figure S11c,d) can also be obtained at 293 K (Tetragonal phase).

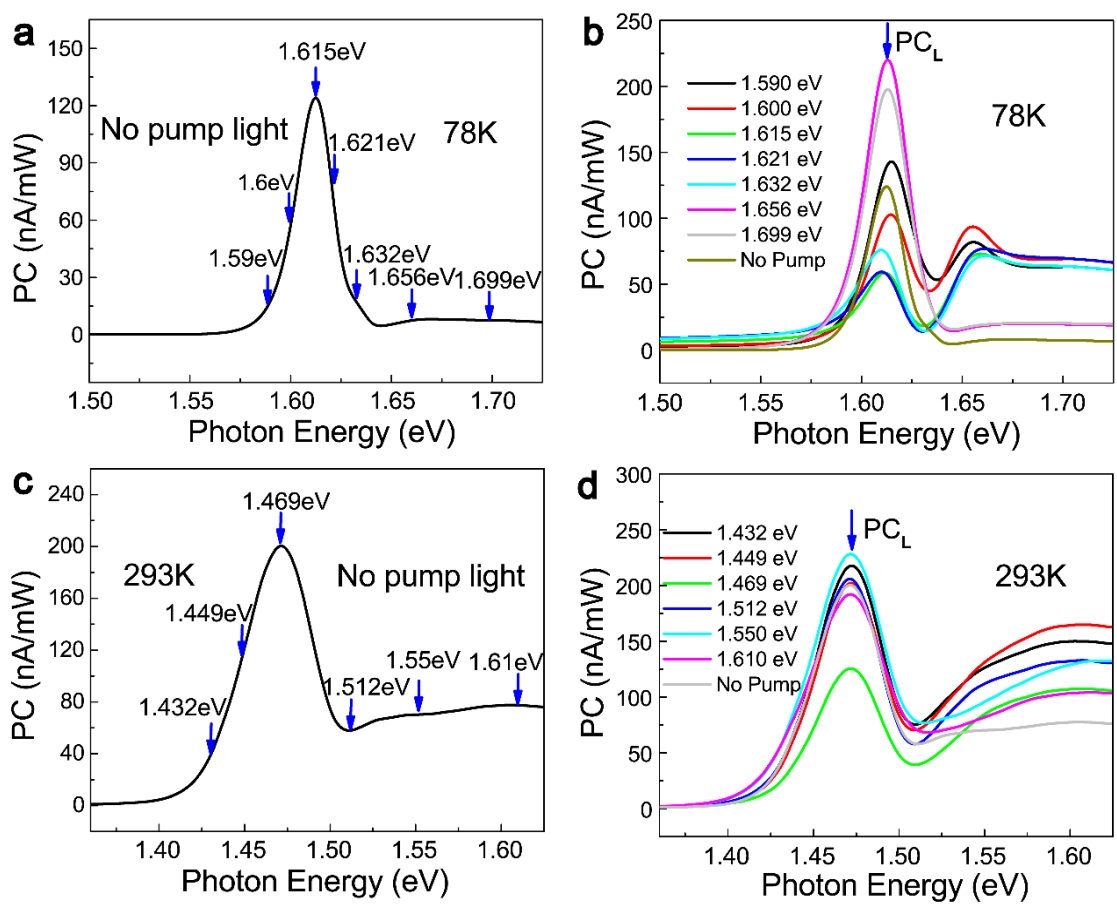

**Figure S11.** The Pump-probe PC spectra of the MSC. (a) The PC spectrum of the sample without pump light at 78 K. (b) The PC spectrum induced by different pump lights at 78 K. (c) The PC spectrum of the sample without pump light at 293 K. (d) The PC spectrum induced by different pump lights at 293 K.

**Table S1.** The extracted fitting values of the linewidth parameters of $PL_H$ and $PC_L$. Linewidth broadening parameters extracted from fits of $\Gamma(T) = \Gamma_0 + \gamma_{LO}/(e^{E_{LO}/K_BT} - 1)$ to the $PL_H$ and $PC_L$ linewidth data. $\Gamma_0$ is the inhomogeneous broadening (the linewidth at 0 K), $\gamma_{LO}$ is the strength of the LO phonon-charge-carrier Fröhlich coupling and $E_{LO}$ is the relevant LO phonon energy.

| Fitting Parameters | High-energy Transition ($PL_H$) | Low-energy Transition ($PC_L$) |
|---|---|---|
| $\Gamma_0$ (meV) | 20 | 18 |
| $\gamma_{LO}$ (meV) | 200 | 120 |
| $E_{LO}$ (meV) | 25 | 40 |